\documentclass[preprints,article,accept,moreauthors,pdftex]{Definitions/mdpi} 
\firstpage{1} 
\pubvolume{1}
\issuenum{1}
\articlenumber{0}
\pubyear{2021}
\copyrightyear{2021}
\datereceived{} 
\dateaccepted{} 
\datepublished{} 
\hreflink{https://doi.org/} 
\pdfoutput=1

\newcommand{\code}[1]{\texttt{#1}}

\Title{Turning AGN bubbles into radio relics with sloshing: modeling CR transport with realistic physics}

\TitleCitation{Turning AGN bubbles into radio relics with sloshing: modeling CR transport with realistic physics}

\Author{John ZuHone $^{1,*}$\orcidA{}, Kristian Ehlert $^{2}$\orcidB{}, Rainer Weinberger $^{3}$\orcidC{} and Christoph Pfrommer $^{2}$\orcidD{}}

\AuthorNames{John ZuHone, Kristian Ehlert, Rainer Weinberger and Christoph Pfrommer}

\AuthorCitation{ZuHone, J.; Ehlert, K.; Weinberger, R.; Pfrommer, C.}

\address{%
$^{1}$ \quad Center for Astrophysics $\vert$~Harvard \& Smithsonian, 60 Garden St., Cambridge, MA 02138, USA\\
$^{2}$ \quad Leibniz-Institut f\"ur Astrophysik Potsdam (AIP), An der Sternwarte 16, 14482 Potsdam, Germany\\
$^{3}$ \quad Canadian Institute for Theoretical Astrophysics, 60 St. George St., Toronto, ON M5S 3H8, Canada\\
}

\corres{Correspondence: john.zuhone@cfa.harvard.edu}

\abstract{Radio relics are arc-like synchrotron sources at the periphery of galaxy clusters, produced by cosmic-ray electrons in a $\mu$G magnetic field which are believed to have been (re-)accelerated by merger shock fronts. However, not all relics appear at the same location as shocks as seen in the X-ray. In a previous work, we suggested that the shape of some relics may result from the pre-existing spatial distribution of cosmic-ray electrons, and tested this hypothesis using simulations by launching AGN jets into a cluster atmosphere with sloshing gas motions generated by a previous merger event. We showed that these motions could transport the cosmic ray-enriched material of the AGN bubbles to large radii and stretch it in a tangential direction, producing a filamentary shape resembling a radio relic. In this work, we improve our physical description for the cosmic rays by modeling them as a separate fluid which undergoes diffusion and Alfv\'en losses. We find that including this additional cosmic ray physics significantly diminishes the appearance of these filamentary features, showing that our original hypothesis is sensitive to the modeling of cosmic ray physics in the intracluster medium.}


\begin{document}

\end{paracol}

\section{Introduction}\label{sec:intro}

In galaxy cluster mergers, the kinetic energy of the clusters' hot plasma (the intracluster medium,
ICM) is dissipated via shock waves and turbulence into heat, amplification of magnetic fields, and
acceleration of cosmic rays (CRs). CR electrons (CRe) with $\gamma \sim 10^3-10^4$ are observed
in clusters via synchrotron radio emission in the $\sim\mu$G magnetic field of clusters. The
brightest sources are the radio lobes associated with active galactic nuclei (AGN), produced by jets
from black holes (BHs). 

However, many clusters have other radio sources which are more diffuse and are not directly
associated with an AGN. These sources generally fall into two categories: radio halos and radio
relics \cite{vanWeeren2019}. Radio halos are volume-filling, unpolarized, and diffuse, and can either
be ``giant'' or ``mini'' radio halos, the former occuring in clusters with major mergers and the
latter seen in the cool cores of relaxed clusters. Radio halos are thought to arise from the
reacceleration of CRe with $\gamma \sim 10^2$ by merger-driven turbulence
\cite{Brunetti2007,Brunetti2014,Pinzke2017}, but radio mini-halos may originate via turbulence \cite{ZuHone2013} or
hadronic interactions between cosmic ray protons (CRp) and the thermal population, producing
secondary CRe which emit in the radio \cite{Pfrommer2004a,Pfrommer2004b,ZuHone2015}. Radio relics
(or ``radio gischts'' \cite{Kempner2004}, ``radio shocks'' \cite{vanWeeren2019}) are typically found
in the outskirts of clusters, are elongated, and strongly polarized. These narrow radio arcs are
usually concentric with the cluster X-ray emission and sometimes have counterparts on the opposite
side of the cluster. 

Originally it was assumed that these latter features mark the location of shock fronts propagating
in the ICM which accelerate ICM electrons to ultrarelativistic energies via a first-order Fermi
process, which would then cool rapidly and produce narrow ($\sim$100 kpc) radio features. It has
been shown that this scenario implies an implausibly high acceleration efficiency for the low-Mach
(${\cal M} \lesssim 3$) cluster merger shocks \cite[e.g.,][]{Macario2011}. It is more likely that
shocks re-accelerate long-lived CRe with lower $\gamma \sim 10^2$ from past acceleration events
\cite{Sarazin1999,Markevitch2005,Pinzke2013,Kang2016}. Aside from this, not all radio relics exhibit ICM shocks
in the X-ray at their front edges, as expected in a simple re-acceleration picture, the best example
of this being the most prominent and well-studied CIZA\,J2242.8+5301 Sausage relic
\cite{Ogrean2013,Ogrean2014,vanWeeren2010}. 

In a previous work \cite[][hereafter Z21]{ZuHone2021}, we explored the alternative possibility that
radio relics such as the Sausage trace the nonuniform distribution of the seed CRe rather than the
ICM shocks. Assuming the distribution of CRe originates from injections by AGN from the cluster
center and other galaxies which are subsequently transported throughout the cluster by gas motions,
we may expect that those regions have sharp boundaries and shapes as long arched filaments or sheets
concentric with the cluster, and that the magnetic fields within those regions are stretched along
the filaments. A shock passage across such a region would create a radio relic with all the observed
relic properties. We may catch the shock front while it crosses such a region, creating a radio
relic at the shock position. When the shock moves out (but not too far, in order to satisfy the
electron cooling timescale constraint), the relic ``front edge'' would delineate the sharp boundary
of the polluted region. 
 
In Z21, we tested this hypothesis using magnetohydrodynamic (MHD) simulations of AGN feedback in
clusters which have had their ICM stirred by recent merging activity. In these simulations, the
central BH fires jets into the cluster core, producing highly magnetized buoyant bubbles which then
rise. As they rise, the material from the bubbles is advected by the merger-driven gas motions, and
eventually this material is mixed with the ICM. We were able to track the evolution of the material
from the bubbles by following a passive tracer injected with the jet and advected with the gas. We
showed that in some cases, the gas motions can be fast and widespread enough to advect the bubble
material from the cluster core to very large radii, and produce large, elongated ($\sim$1~Mpc),
cluster-centric CR-enriched regions that resemble radio relics. We also showed that in these
regions, magnetic fields are naturally stretched tangentially along the filaments, helping to
explain the large degree of polarization in relics.

In Z21, the physics of the injected jet material was very simplified--the same kind of magnetized,
thermal plasma as the ICM itself. The ``cosmic rays'' in the simulations were marked by a passive
scalar injected by the jet, which is subsequently advected by gas motions. However, the physics of
the CRs which are injected into ICM by jets is more complicated than this in reality. These CRs are unstable to the streaming instability \cite{Kulsrud1969}
and resonantly excite Alfv\'en waves, on which they subsequently scatter and transfer their energy
into heat. Depending on the prevailing damping process of these Alfv\'en waves, CRs can either stream (if damping is weak), diffuse (if damping is strong), or are transported by a combination of both processes along the direction of the local
magnetic field \cite{Wiener2013}.

In this work, we implement this additional CR physics in two of the simulations from Z21, to determine if our previous results regarding the formation of relic-like CRe features are affected. 

\section{Methods}\label{sec:methods}

The simulations we perform consist of 1) an idealized binary merger between two clusters with a high
mass ratio, in which a small subcluster perturbs the center of a large, cool-core cluster,
initiating sloshing motions and turbulence, followed by 2) the injection of AGN jets from a central
BH. Since most of the details of the simulations have been discussed in Z21, we will only
briefly outline them here, and devote more discussion to the additional physics added in this work. 

We perform our simulations using the \code{AREPO} code \citep[][]{Springel2010}, which employs a
finite-volume Godunov method on an unstructured moving Voronoi mesh to evolve the equations of
magnetohydrodynamics (MHD), and a Tree-PM solver to compute the self-gravity from gas and dark
matter. The magnetic fields are evolved on the moving mesh using the Powell 8-wave scheme with
divergence cleaning employed in \citep{Pakmor2013} and in the IllustrisTNG simulations
\citep{Marinacci2018}. These simulations also include dark matter particles, which make up the bulk
of the cluster's mass and only interact with each other and the gas via gravity. 

\subsection{Merger Simulations}\label{sec:merger}

The idealized binary merger simulation consists of a large, ``main'' cluster with a total mass of $M \sim 10^{15}~M_\odot$, and a smaller infalling subcluster five times less massive. The main cluster includes DM and hot plasma, where the fraction of gas mass to dark matter mass is 0.12. The subcluster is DM-only (perhaps stripped of its gas on a previous cluster passage). The initial conditions in the simulation explored here are the same as in previous works \cite[][and Z21]{Ascasibar2006,ZuHone2010,ZuHone2016,ZuHone2018,ZuHone2019}, and produce spiral-shaped cold fronts in an otherwise relatively relaxed cluster very similar to those observed in X-ray observations. In Z21, this simulation also produced thin strips of bubble material which closely resembled radio relics more than the ``Merger2'' simulation from that work.

The plasma is modeled as a magnetized, fully ionized ideal fluid with $\gamma = 5/3$ and mean molecular weight $\mu = 0.6$. For the magnetic field in the
simulation, we follow the procedure from \cite{Brzycki2019} and references therein. Briefly, we set
up a divergence-free turbulent magnetic field on a uniform grid with a Kolmogorov spectrum which is
isotropic in the three spatial directions. The average magnetic field strength is scaled to be
proportional to the square root of the thermal pressure such that $\beta = P_{\rm th}/P_B$ is
constant, with an initial value of $\beta = 100$ everywhere. The magnetic field components are then interpolated from this grid onto the cells in the simulation.

\subsection{Simulation of Jets and Cosmic Rays}\label{sec:jets_crs}

For simulating the effects of AGN in the simulations we use the method of \cite{Weinberger2017}.
This method injects a bi-directional jet from a BH particle located at the cluster potential
minimum, which is kinetically dominated, low density, and collimated. Kinetic, thermal, and magnetic
energy is injected into two small spherical regions a few kpc from the location of a black hole
particle. The material injected by the jet is marked by a passive tracer field $X_{\rm jet}$ and is
advected along with the fluid for the rest of the simulation. 


\begin{figure}[t]
\widefigure
\includegraphics[width=0.9\textwidth]{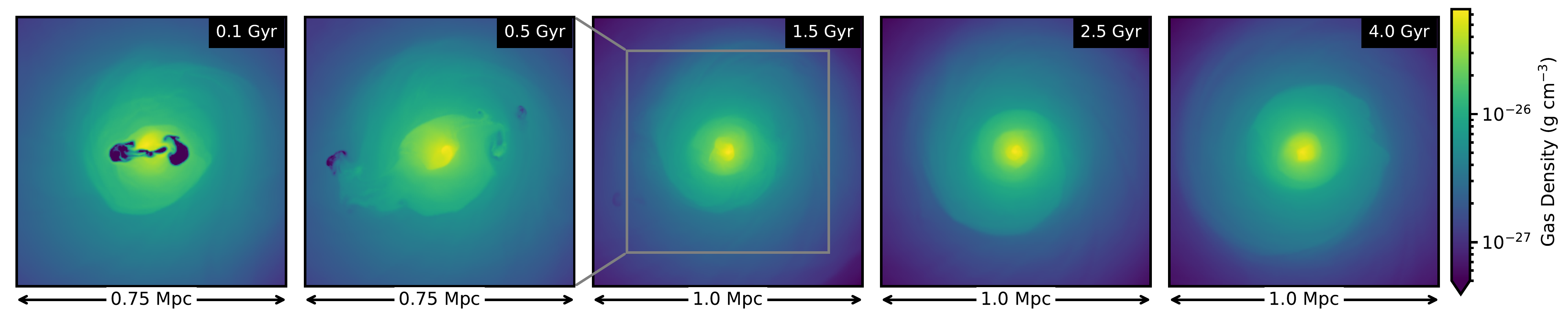}
\caption{Slices through the cluster center of the gas density of the ``Merger1''/jet simulation from Z21 with the jets oriented along the $x$-axis. Several epochs are shown with time $\tau$
measured from the moment of jet injection.\label{density}}
\end{figure}   

            
In each simulation, the jets are fired with a luminosity $L_{\rm jet} = 3.169 \times
10^{45}$~erg~s$^{-1}$ for a duration of $t_{\rm jet} = 100$~Myr, so the resulting total energy
injected is $E_{\rm jet} = 10^{61}$~erg in each direction, which is a sum of kinetic, thermal, and
magnetic energy. In the jet region, the magnetic and thermal pressures are equal ($\beta_{\rm jet} =
P_{\rm th}/P_B = 1$), and the injected magnetic field is purely toroidal. As in Z21, we fire the jet
once and follow the subsequent evolution of the material it injects into the ICM.

In Z21, the jet was comprised purely of thermal gas and magnetic energy, and the only marker for the
material injected by the jet was the passive tracer field $X_{\rm jet}$. In this work, we improve
upon the physical description of the jet material by following \cite{Pfrommer2017,Ehlert2018} in
treating the CRs as a secondary fluid with $\gamma_{\rm CR} = 4/3$. We set the CR and thermal
pressures in the jet to be equal ($\beta_{\rm CR, jet} = P_{\rm th}/P_{\rm CR} = 1$), so that the
total pressure in the jet is equal parts from thermal plasma, CRs, and magnetic fields. We emulate a
simple model of CR acceleration by internal shocks within the jet \cite{Perucho2007} by maintaining
$\beta_{\rm CR, jet} = 1$ for $t_{\rm acc} = 2t_{\rm jet}$ within cells which have $X_{\rm jet} >
10^{-3}$. 
    
CR diffusion and streaming is emulated via a combination of diffusion and Alfv\'en losses \citep{Wiener2017}. The CR energy density $\varepsilon_\mathrm{CR}$ is advected with the gas and anisotropically diffuses along the magnetic field lines, with a constant diffusion coefficient $\kappa_\parallel$ along the local magnetic field
direction and $\kappa_\perp = 0$ perpendicular to it \cite{Pakmor2016}. 
For more details of the CR algorithms used here see \cite{Ehlert2018} and references therein.
We compare three models: 1) the original model of Z21 without CRs, 2) the same cluster and AGN setup but with purely advective CRs (without diffusion), and 3) a model with a realistic CR diffusion coefficient in the ICM of $\kappa_\parallel = 10^{29}$~cm$^2$~s$^{-1}$. The cross-field diffusion is significantly smaller in comparison to diffusion along the magnetic field \cite[e.g.][]{Desiati2014} and thus our numerical cross-field diffusion of cosmic rays likely dominates over the physical transport in the direction perpendicular to the magnetic field \cite{Pakmor2016}. We demonstrate that the resulting cosmic ray distribution is mainly shaped by advective cosmic ray transport as a result of the sloshing ICM over diffusive transport along the magnetic field or our assumed diffusion coefficient of $10^{29}$~cm$^2$~s$^{-1}$, which is motivated by the cosmic ray streaming picture \cite{Ehlert2018}.
            
In model 3, The effective diffusion velocity across a CR gradient length of $L_\mathrm{CR}=\varepsilon_\mathrm{CR}/\nabla \varepsilon_\mathrm{CR} = 3$~kpc is $v_{\mathrm{diff}}\sim\kappa_\parallel /L_\mathrm{CR}\sim 100$~km~s$^{-1}$. Thus, CRs are expected to fill a significant volume of the cluster core. However, the sloshing front is expected to stretch the magnetic field lines and approximately align them with the equipotential surfaces. The CR gradient length along the magnetic field lines increases substantially so that the diffusion velocity decreases and thereby confines CRs within the sloshing front. 
            
    
\section{Results}\label{sec:results}
     
\begin{figure}[t]
\centering
\includegraphics[width=0.9\textwidth]{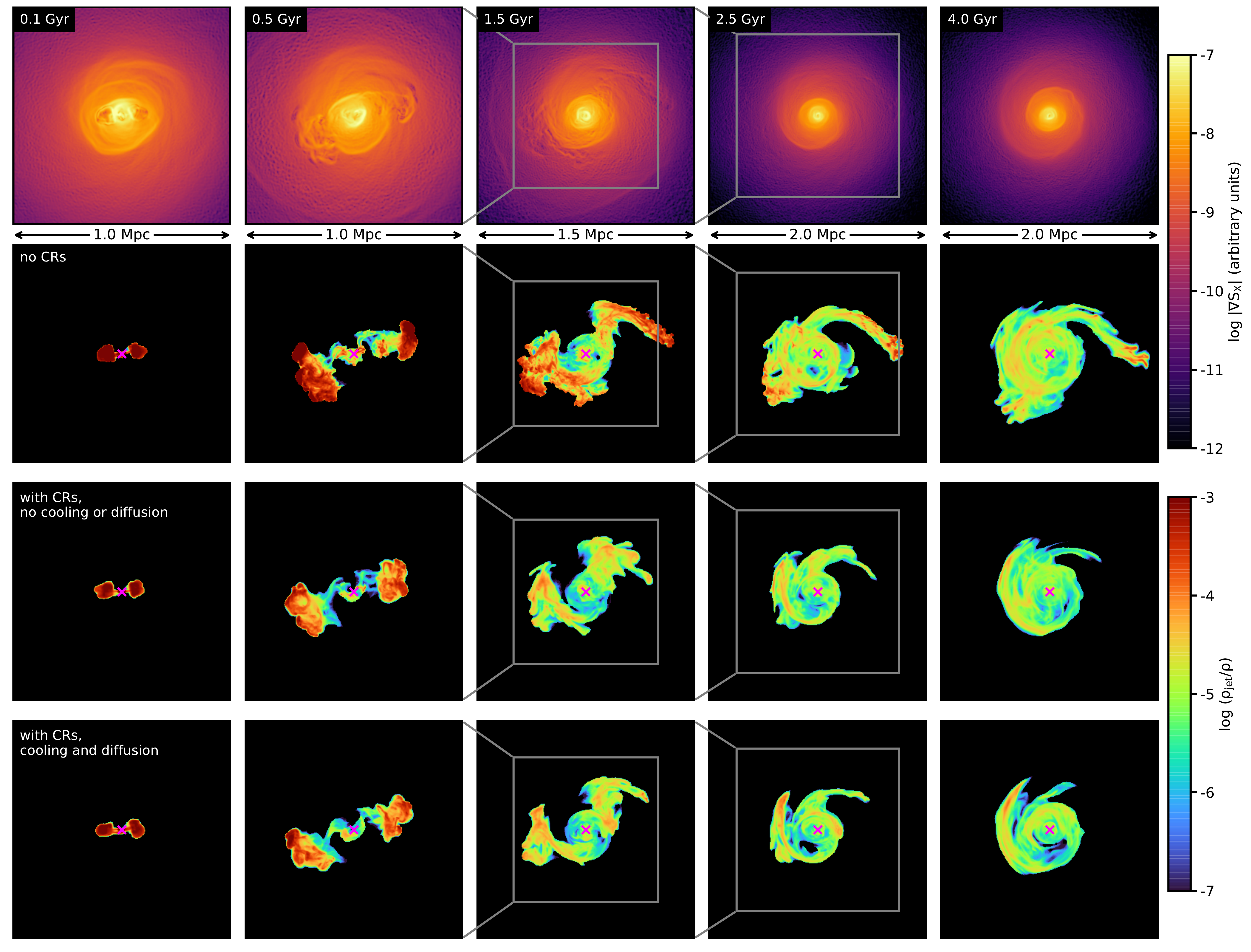}
\caption{Projections along the $z$-axis of the ``Merger1''/jet simulation from Z21 with the jets
oriented along the $x$-axis for varying CR physics. Several epochs are shown with time $\tau$
measured from the moment of jet injection. (\textbf{a}) The top row shows the X-ray SB gradient
magnitude. (\textbf{b}) The second row shows the mass-weighted passive tracer $X_{\rm jet}$ from the
original Z21 simulation, without additional CR physics. (\textbf{c}) The third row shows the same
simulation, now with the CRs included, but these are only advected with the fluid. (\textbf{d}) The fourth row shows the same simulation as in the third, but in this case diffusion and Alfv\'en losses are included. \label{jets_x}}
\end{figure}   
    
Here we present the results of the ``Merger1''/jet simulations from Z21 with jets oriented along the
$x$-axis of the simulation box with additional CR physics included. These are shown in
Figures~\ref{density} and \ref{jets_x}. The simulation epoch $\tau$ of the quantity plotted in each row increases from left to right, with $\tau = 0$ corresponding to the moment of jet injection. Figure~\ref{density} shows slices through the gas density from one of the simulations, showing the initial injection of the bubbles and their subsequent mixing into the ICM, followed by the continuation of the sloshing motions. The top row in Figure~\ref{jets_x} shows the gradient magnitude of the projected X-ray surface brightness (SB) along the $z$-axis of the simulation box (from the simulation without CR physics, though the differences in the simulations are negligible in terms of the effect on the sloshing gas), to serve as a visual aid to locate the position of the cold fronts and the sloshing motions which are advecting the CRs.
    
The following rows of panels in Figure~\ref{jets_x} shows the mass-weighted projected passive tracer
field $X_{\rm jet}$ along the $z$-axis, which is common to all simulations. The second row shows
projected jet material in the original simulation from Z21, without CRs included in the jets. The
jet-injected bubbles rise, and as their material is mixed with the surrounding ICM, it follows the
sloshing gas motions. The inner region bounded by the cold fronts is filled with material originally
injected by the jet, and a $\sim$Mpc-long strip of jet material appear at large radius by $\tau \sim
4.0$~Gyr (rightmost panels of Figure~\ref{jets_x}), to the NW of the cluster core.      

The third row shows projected jet material in the simulation where only the CR component has been added to the original simulation of Z21, without any additional physics. Here, the distribution of projected jet material looks qualitatively similar to the original simulation for the first $\tau = 1.5$~Gyr, but after this it is obvious that the bubble to the W has not risen to the same height, and is unable to get caught up in the bulk motions to the N/NW to get stretched into a large filament.

The fourth row of Figure~\ref{jets_x} shows the simulation where diffusion and Alfv\'en losses for the CRs have been turned on. These look qualitatively very similar to the simulation with purely advective CRs (model 2), but the smoothing out of CR gradients by diffusion has had a noticeable effect on the distribution of jet material which closely tracks the CRs in this simulation.


\begin{figure}
\widefigure
\includegraphics[width=0.95\textwidth]{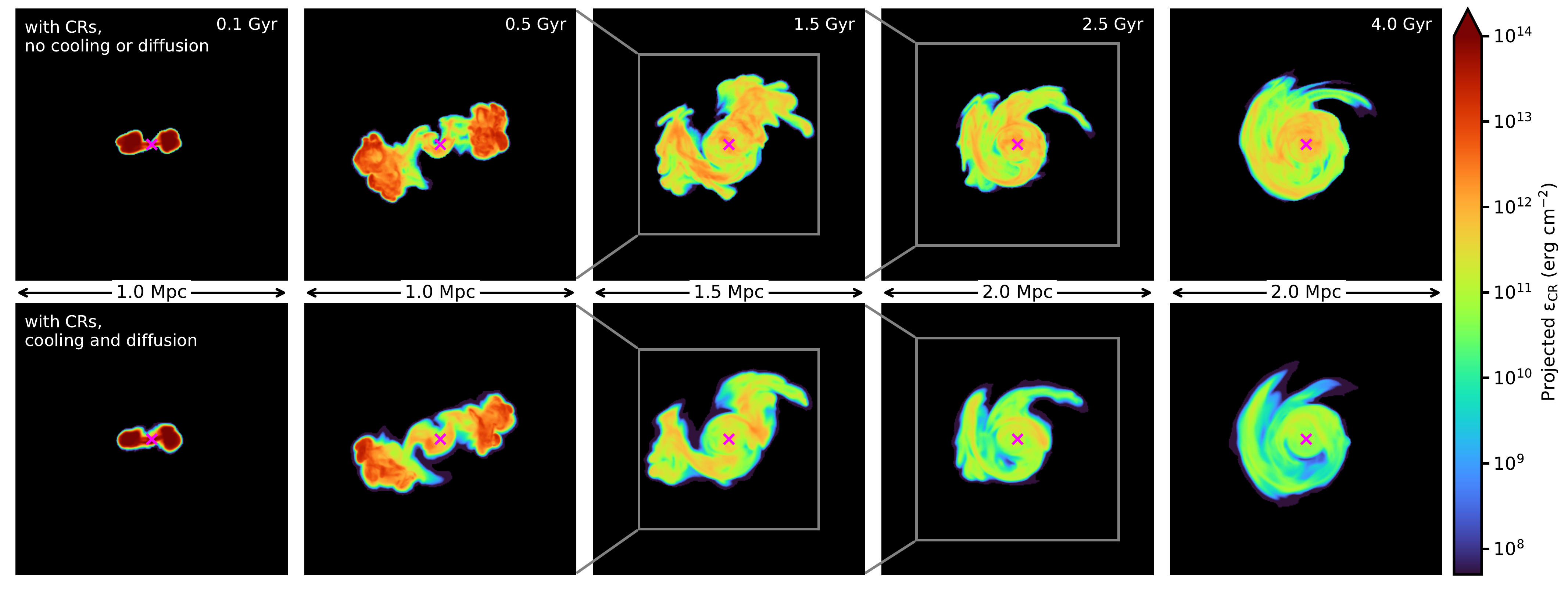}
\caption{Projected CR energy density along the $z$-axis in the same simulations with jets
fired in the $x$-direction, for the two simulations with CRs included, with and without the effects of diffusion and Alfv\'en losses. Several epochs are shown with time $\tau$ measured from the moment of jet injection.\label{cr_density}}
\end{figure}   

    
Figure \ref{cr_density} shows the projected CR energy density in the same simulations. Overall, the
projected CR distribution matches the jet tracer field distribution. However, these maps more
clearly show the effect of diffusion and Alfv\'en losses; when diffusion and cooling are present,
the CR distribution is smoother, with a less irregular appearance than in the case with no cooling
or diffusion, which merely support advective CR transport. Alfv\'en losses have reduced the CR
energy density by roughly an order of magnitude. Diffusion weakens the gradients
along a magnetic field line while Alfv\'en losses dissipate CR energy at a rate
$|\boldsymbol{v}_\mathrm{A}\boldsymbol{\cdot\nabla} P_\mathrm{CR}|$ (where
$\boldsymbol{v}_\mathrm{A}$ denotes the Alfvén velocity) and thus sharpens the CR gradients. As a
result CR gradients are maintained along filamentary magnetic field lines at the edges of our CR
distribution towards the outer cluster regions. This effectively precludes the formation of such
long filamentary features in the CR energy density. On the opposite side, within the core spiral
structure, CR diffusion homogenizes the CR energy density, softens CR gradients, and thus minimizes
Alfv\'en losses.

In Figure \ref{mag_vectors} we show the projected CR energy density along the $z$-axis of the
simulations with CRs at $\tau = 3.5$~Gyr, with plane-of-sky projected magnetic field vectors
overlaid. The projected magnetic field has been weighted by a factor $B^2\varepsilon_{\rm CR}$,
which will be approximately proportional to radio emission from secondary CR electrons that are
produced in hadronic CR interactions with the ambient ICM emissivity $I(\nu) \propto \nu^{-\alpha}$
with $\alpha = 1$. In Z21, elongated regions of enhanced CR distribution were associated with
magnetic field lines which are largely aligned lengthwise along these features, which naturally
arise due to the fact that the same motions which spread the CRs out over long distances stretch and
amplify the magnetic field along the same direction. While this is also generally true in the
simulations presented here with CRs, we note that in the case with cooling and diffusion (right
panel of Figure \ref{mag_vectors}) the degree of alignment between the field line direction and the
long axis of the filamentary features appears somewhat degraded in those features which are left,
especially for the one $\sim$300~kpc north of the cluster center, indicating that these processes have altered the distribution of CRs such that the ones that remain are no longer in regions where the magnetic field is tangentially aligned as seen in projection.

\section{Discussion and Conclusions}\label{sec:conc}

Z21 showed that merger-driven motions in the ICM could advect material injected by AGN jets to large
radii and produce Mpc-scale thin filaments of jet material, which were posited to explain the
appearance of radio relics that are not explained by the simple model of a merger shock
overunning a region uniformly filled with CRe in the cluster outskirts. Under this hypothesis, the
appearance of these relics emerges due to the spatial distribution of the CRe themselves as
determined by the ICM motions which transported and stretched them across large distances. We have
shown in this work that for a more realistic model of CRs in the simulation, whether or not we include the effects of CR diffusion and Alfv\'en losses, filamentary features become less prominent, and are unable to reach the same large radii that they were in the
absence of the additional CR physics. 

What accounts for this difference? One obvious source of this behavior is the difference in adiabatic index between a bubble which is filled with thermal gas and a bubble with an energy density which is half CRs. The reduced pressure can destabilize the rising bubble. The CR acceleration process itself may also effect the stability of the jet. CRs have a slower cooling time in comparison to the thermal gas in the center of cool cores so that they can prevent runaway thermal cooling and locally provide thermal stability \cite{Pfrommer2013}, although gas in collisional ionization equilibrium can become unstable for certain conditions \cite{Kempski2020}. All of these issues will be examined in more detail in future simulations of jets combined with merger-driven gas motions. 

For the simulation with CR streaming, Alfv\'en wave losses will also preclude the formation of extended filaments by dissipating the CR energy density. The formation of filaments also necessarily depends on the precise details of the cluster ``weather'' that a given bubble is encountering. A wider parameter space of jet axes, jet energies, and cluster gas motion characteristics is necessary to determine if the hypothesis suggested by Z21 regard a possible origin for radio relics holds up under the assumption of more realistic models for CR dynamics.


\begin{figure}
\widefigure
\includegraphics[width=0.48\textwidth]{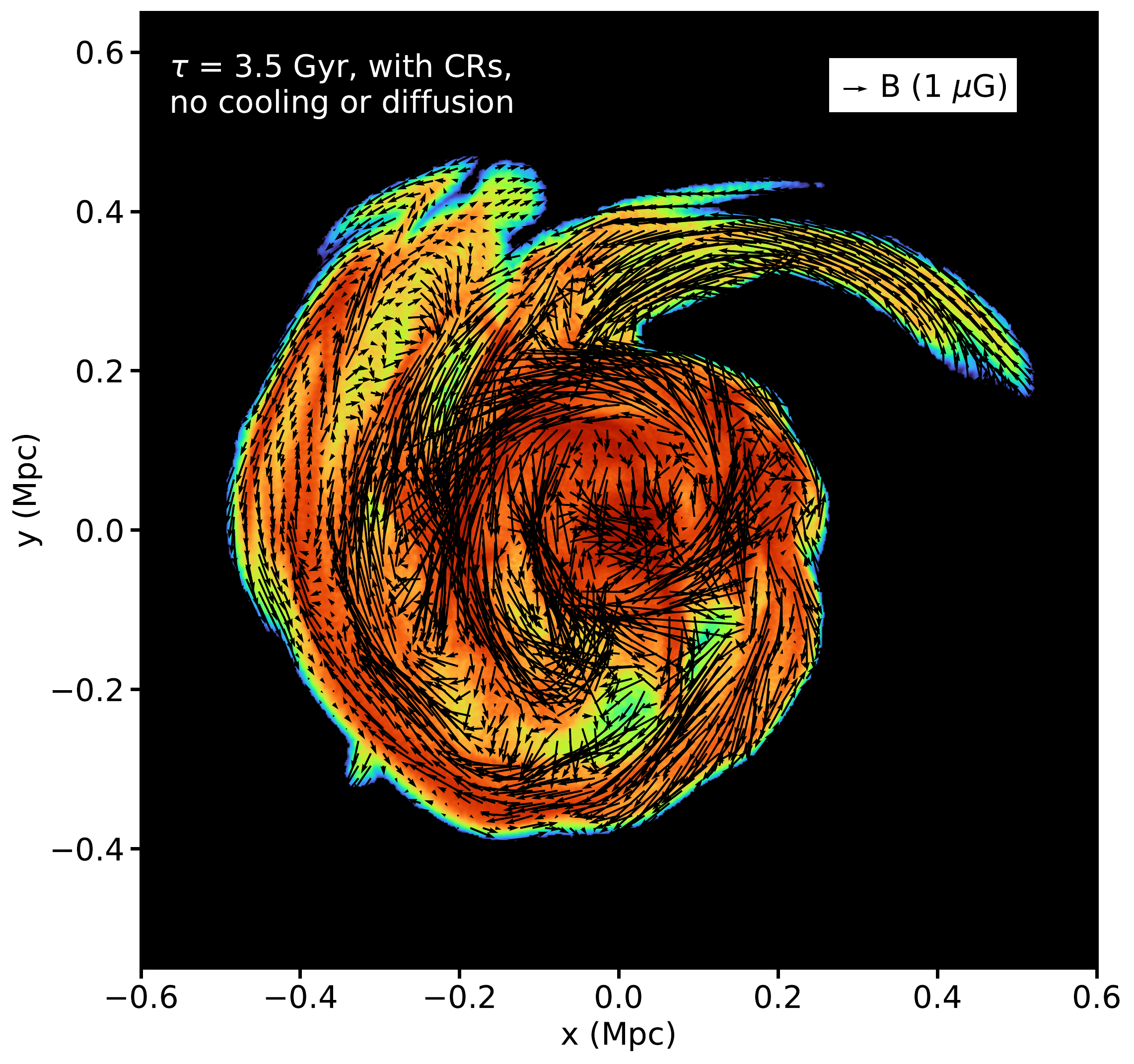}
\includegraphics[width=0.48\textwidth]{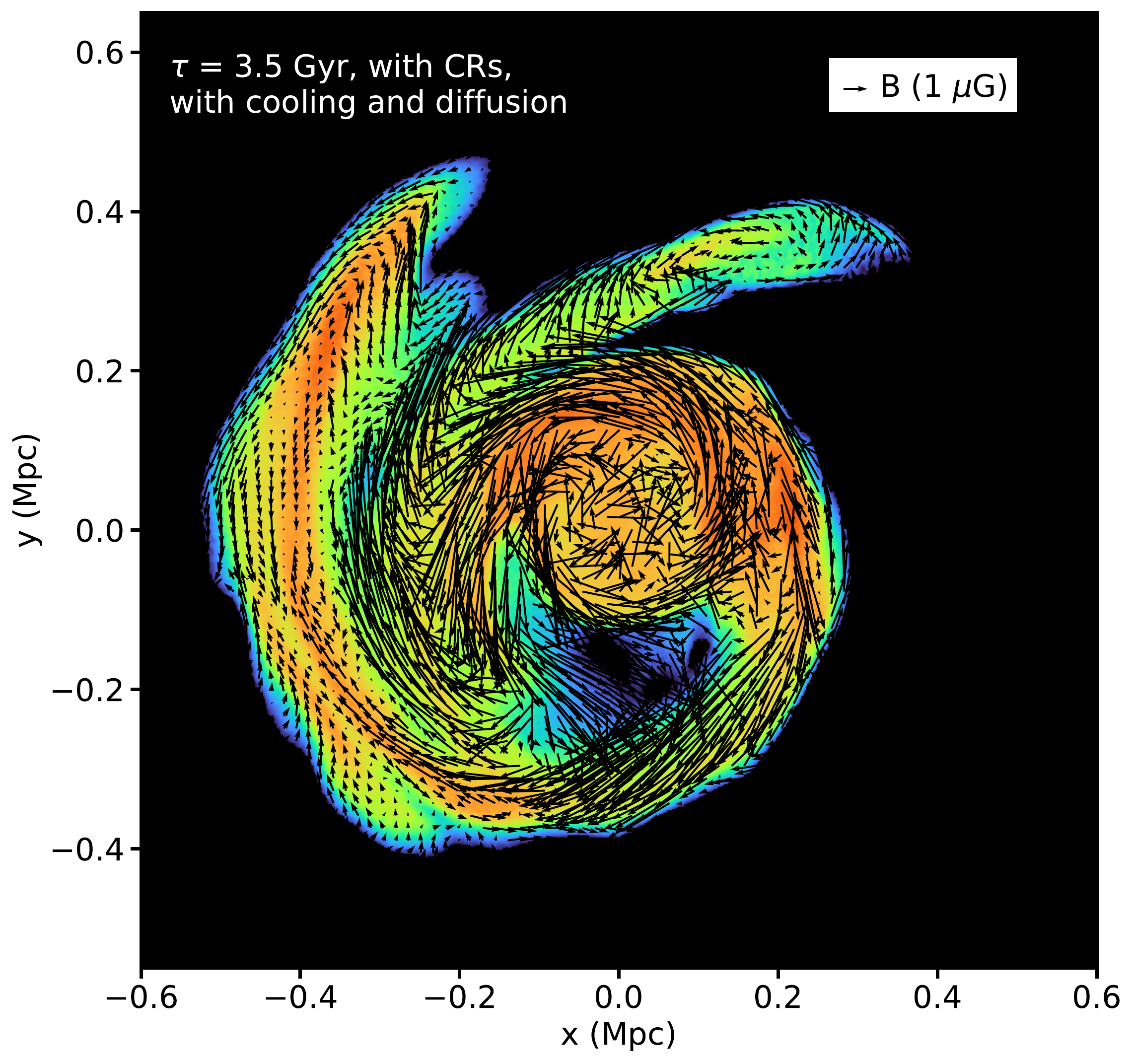}
\caption{Projected CR energy density and plane-of-sky magnetic field vectors for the same
simulations. The magnetic field vectors are weighted proportional to $B^2\varepsilon_{\rm
CR}$.\label{mag_vectors}}
\end{figure}   

    
Also, our simulations have only included one jet phase of the central AGN. In reality the CR
distribution in a cool-core cluster will be the superposition of many such injections, which will
have been transported around the core region by gas motions in different stages. Future work should
include a self-consistent feedback model which is fed by gas which is radiative cooling to determine
what the expected long-term distribution of CRs in a sloshing cool core would be. We also observe that diffusion appears to isotropize the CRs confined within the core region (right panels of Figures \ref{cr_density} and \ref{mag_vectors}), suggesting that this may provide an explanation for the existence of radio mini-halos in some clusters via emission from secondary electrons \cite{Pfrommer2004a,Pfrommer2004b}. 

\vspace{6pt} 



\authorcontributions{Conceptualization, J.Z. and C.P.; methodology, J.Z., K.E., and R.W.; software, K.E., R.W.; validation, J.Z.; formal analysis, J.Z.; investigation, J.Z.; resources, J.Z.; data curation, J.Z.; writing---original draft preparation, J.Z.; writing---review and editing, J.Z., K.E., R.W., and C.P.; visualization, J.Z.; funding acquisition, J.Z. All authors have read and agreed to the published version of the manuscript.}

\funding{This research was funded by the Chandra X-ray Center, which is operated by the Smithsonian Astrophysical Observatory for and on behalf of NASA under contract NAS8-03060.}

\dataavailability{Simulation data will be made available at the Galaxy Cluster Merger Catalog\footnote{\url{http://gcmc.hub.yt}}\cite{ZuHoneGCMC}, or by request to the author.} 

\acknowledgments{We thank Maxim Markevitch for useful discussions. The simulations were run on the Pleiades supercomputer at NASA/Ames Research Center. Software packages used in this work include: AREPO\footnote{\url{https://arepo-code.org}} \cite{Springel2010,Weinberger2017}; AstroPy\footnote{\url{https://www.astropy.org}} \cite{AstroPy2013};
Matplotlib\footnote{\url{https://matplotlib.org}} \cite{Hunter2007};
NumPy\footnote{\url{https://www.numpy.org}} \cite{Harris2020};
yt\footnote{\url{https://yt-project.org}} \cite{Turk2011}}

\conflictsofinterest{The authors declare no conflict of interest.} 


\reftitle{References}


\externalbibliography{yes}
\bibliography{main.bib}

\begin{thebibliography}{999}

\bibitem[{van Weeren} \em{et~al.}(2019){van Weeren}, {de Gasperin}, {Akamatsu},
  {Br{\"u}ggen}, {Feretti}, {Kang}, {Stroe}, and {Zandanel}]{vanWeeren2019}
{van Weeren}, R.J.; {de Gasperin}, F.; {Akamatsu}, H.; {Br{\"u}ggen}, M.;
  {Feretti}, L.; {Kang}, H.; {Stroe}, A.; {Zandanel}, F.
\newblock {Diffuse Radio Emission from Galaxy Clusters}.
\newblock {\em Space Science Reviews} {\bf 2019}, {\em 215},~16,
  \href{http://xxx.lanl.gov/abs/1901.04496}{{\normalfont
  [arXiv:astro-ph.HE/1901.04496]}}.
\newblock
  doi:{\changeurlcolor{black}\href{https://doi.org/10.1007/s11214-019-0584-z}{\detokenize{10.1007/s11214-019-0584-z}}}.

\bibitem[{Brunetti} and {Lazarian}(2007)]{Brunetti2007}
{Brunetti}, G.; {Lazarian}, A.
\newblock {Compressible turbulence in galaxy clusters: physics and stochastic
  particle re-acceleration}.
\newblock {\em Monthly Notices of the Royal Astronomical Society} {\bf 2007},
  {\em 378},~245--275,
  \href{http://xxx.lanl.gov/abs/astro-ph/0703591}{{\normalfont
  [arXiv:astro-ph/astro-ph/0703591]}}.
\newblock
  doi:{\changeurlcolor{black}\href{https://doi.org/10.1111/j.1365-2966.2007.11771.x}{\detokenize{10.1111/j.1365-2966.2007.11771.x}}}.

\bibitem[{Brunetti} and {Jones}(2014)]{Brunetti2014}
{Brunetti}, G.; {Jones}, T.W.
\newblock {Cosmic Rays in Galaxy Clusters and Their Nonthermal Emission}.
\newblock {\em International Journal of Modern Physics D} {\bf 2014}, {\em
  23},~1430007--98,  \href{http://xxx.lanl.gov/abs/1401.7519}{{\normalfont
  [arXiv:astro-ph.CO/1401.7519]}}.
\newblock
  doi:{\changeurlcolor{black}\href{https://doi.org/10.1142/S0218271814300079}{\detokenize{10.1142/S0218271814300079}}}.

\bibitem[{Pinzke} \em{et~al.}(2017){Pinzke}, {Oh}, and {Pfrommer}]{Pinzke2017}
{Pinzke}, A.; {Oh}, S.P.; {Pfrommer}, C.
\newblock {Turbulence and particle acceleration in giant radio haloes: the
  origin of seed electrons}.
\newblock {\em Monthly Notices of the Royal Astronomical Society} {\bf 2017},
  {\em 465},~4800--4816,
  \href{http://xxx.lanl.gov/abs/1611.07533}{{\normalfont
  [arXiv:astro-ph.HE/1611.07533]}}.
\newblock
  doi:{\changeurlcolor{black}\href{https://doi.org/10.1093/mnras/stw3024}{\detokenize{10.1093/mnras/stw3024}}}.

\bibitem[{ZuHone} \em{et~al.}(2013){ZuHone}, {Markevitch}, {Brunetti}, and
  {Giacintucci}]{ZuHone2013}
{ZuHone}, J.A.; {Markevitch}, M.; {Brunetti}, G.; {Giacintucci}, S.
\newblock {Turbulence and Radio Mini-halos in the Sloshing Cores of Galaxy
  Clusters}.
\newblock {\em Astrophysical Journal} {\bf 2013}, {\em 762},~78,
  \href{http://xxx.lanl.gov/abs/1203.2994}{{\normalfont
  [arXiv:astro-ph.CO/1203.2994]}}.
\newblock
  doi:{\changeurlcolor{black}\href{https://doi.org/10.1088/0004-637X/762/2/78}{\detokenize{10.1088/0004-637X/762/2/78}}}.

\bibitem[{Pfrommer} and {En{\ss}lin}(2004{\natexlab{a}})]{Pfrommer2004a}
{Pfrommer}, C.; {En{\ss}lin}, T.A.
\newblock {Constraining the population of cosmic ray protons in cooling flow
  clusters with {\ensuremath{\gamma}}-ray and radio observations: Are radio
  mini-halos of hadronic origin?}
\newblock {\em Astronomy \& Astrophysics} {\bf 2004}, {\em 413},~17--36.
\newblock
  doi:{\changeurlcolor{black}\href{https://doi.org/10.1051/0004-6361:20031464}{\detokenize{10.1051/0004-6361:20031464}}}.

\bibitem[{Pfrommer} and {En{\ss}lin}(2004{\natexlab{b}})]{Pfrommer2004b}
{Pfrommer}, C.; {En{\ss}lin}, T.A.
\newblock {Estimating galaxy cluster magnetic fields by the classical and
  hadronic minimum energy criterion}.
\newblock {\em Monthly Notices of the Royal Astronomical Society} {\bf 2004},
  {\em 352},~76--90,
  \href{http://xxx.lanl.gov/abs/astro-ph/0404119}{{\normalfont
  [arXiv:astro-ph/astro-ph/0404119]}}.
\newblock
  doi:{\changeurlcolor{black}\href{https://doi.org/10.1111/j.1365-2966.2004.07900.x}{\detokenize{10.1111/j.1365-2966.2004.07900.x}}}.

\bibitem[{ZuHone} \em{et~al.}(2015){ZuHone}, {Brunetti}, {Giacintucci}, and
  {Markevitch}]{ZuHone2015}
{ZuHone}, J.A.; {Brunetti}, G.; {Giacintucci}, S.; {Markevitch}, M.
\newblock {Testing Secondary Models for the Origin of Radio Mini-Halos in
  Galaxy Clusters}.
\newblock {\em Astrophysical Journal} {\bf 2015}, {\em 801},~146,
  \href{http://xxx.lanl.gov/abs/1403.6743}{{\normalfont
  [arXiv:astro-ph.CO/1403.6743]}}.
\newblock
  doi:{\changeurlcolor{black}\href{https://doi.org/10.1088/0004-637X/801/2/146}{\detokenize{10.1088/0004-637X/801/2/146}}}.

\bibitem[{Kempner} \em{et~al.}(2004){Kempner}, {Blanton}, {Clarke},
  {En{\ss}lin}, {Johnston-Hollitt}, and {Rudnick}]{Kempner2004}
{Kempner}, J.C.; {Blanton}, E.L.; {Clarke}, T.E.; {En{\ss}lin}, T.A.;
  {Johnston-Hollitt}, M.; {Rudnick}, L.
\newblock {Conference Note: A Taxonomy of Extended Radio Sources in Clusters of
  Galaxies}.
\newblock  The Riddle of Cooling Flows in Galaxies and Clusters of galaxies;
  {Reiprich}, T.; {Kempner}, J.; {Soker}, N., Eds.,  2004, p. 335,
  \href{http://xxx.lanl.gov/abs/astro-ph/0310263}{{\normalfont
  [arXiv:astro-ph/astro-ph/0310263]}}.

\bibitem[{Macario} \em{et~al.}(2011){Macario}, {Markevitch}, {Giacintucci},
  {Brunetti}, {Venturi}, and {Murray}]{Macario2011}
{Macario}, G.; {Markevitch}, M.; {Giacintucci}, S.; {Brunetti}, G.; {Venturi},
  T.; {Murray}, S.S.
\newblock {A Shock Front in the Merging Galaxy Cluster A754: X-ray and Radio
  Observations}.
\newblock {\em Astrophysical Journal} {\bf 2011}, {\em 728},~82,
  \href{http://xxx.lanl.gov/abs/1010.5209}{{\normalfont
  [arXiv:astro-ph.CO/1010.5209]}}.
\newblock
  doi:{\changeurlcolor{black}\href{https://doi.org/10.1088/0004-637X/728/2/82}{\detokenize{10.1088/0004-637X/728/2/82}}}.

\bibitem[{Sarazin}(1999)]{Sarazin1999}
{Sarazin}, C.L.
\newblock {The Energy Spectrum of Primary Cosmic-Ray Electrons in Clusters of
  Galaxies and Inverse Compton Emission}.
\newblock {\em Astrophysical Journal} {\bf 1999}, {\em 520},~529--547,
  \href{http://xxx.lanl.gov/abs/astro-ph/9901061}{{\normalfont
  [arXiv:astro-ph/astro-ph/9901061]}}.
\newblock
  doi:{\changeurlcolor{black}\href{https://doi.org/10.1086/307501}{\detokenize{10.1086/307501}}}.

\bibitem[{Markevitch} \em{et~al.}(2005){Markevitch}, {Govoni}, {Brunetti}, and
  {Jerius}]{Markevitch2005}
{Markevitch}, M.; {Govoni}, F.; {Brunetti}, G.; {Jerius}, D.
\newblock {Bow Shock and Radio Halo in the Merging Cluster A520}.
\newblock {\em Astrophysical Journal} {\bf 2005}, {\em 627},~733--738,
  \href{http://xxx.lanl.gov/abs/astro-ph/0412451}{{\normalfont
  [arXiv:astro-ph/astro-ph/0412451]}}.
\newblock
  doi:{\changeurlcolor{black}\href{https://doi.org/10.1086/430695}{\detokenize{10.1086/430695}}}.

\bibitem[{Pinzke} \em{et~al.}(2013){Pinzke}, {Oh}, and {Pfrommer}]{Pinzke2013}
{Pinzke}, A.; {Oh}, S.P.; {Pfrommer}, C.
\newblock {Giant radio relics in galaxy clusters: reacceleration of fossil
  relativistic electrons?}
\newblock {\em Monthly Notices of the Royal Astronomical Society} {\bf 2013},
  {\em 435},~1061--1082,  \href{http://xxx.lanl.gov/abs/1301.5644}{{\normalfont
  [arXiv:astro-ph.CO/1301.5644]}}.
\newblock
  doi:{\changeurlcolor{black}\href{https://doi.org/10.1093/mnras/stt1308}{\detokenize{10.1093/mnras/stt1308}}}.

\bibitem[{Kang} and {Ryu}(2016)]{Kang2016}
{Kang}, H.; {Ryu}, D.
\newblock {Re-acceleration Model for Radio Relics with Spectral Curvature}.
\newblock {\em Astrophysical Journal} {\bf 2016}, {\em 823},~13,
  \href{http://xxx.lanl.gov/abs/1602.03278}{{\normalfont
  [arXiv:astro-ph.HE/1602.03278]}}.
\newblock
  doi:{\changeurlcolor{black}\href{https://doi.org/10.3847/0004-637X/823/1/13}{\detokenize{10.3847/0004-637X/823/1/13}}}.

\bibitem[{Ogrean} \em{et~al.}(2013){Ogrean}, {Br{\"u}ggen}, {R{\"o}ttgering},
  {Simionescu}, {Croston}, {van Weeren}, and {Hoeft}]{Ogrean2013}
{Ogrean}, G.A.; {Br{\"u}ggen}, M.; {R{\"o}ttgering}, H.; {Simionescu}, A.;
  {Croston}, J.H.; {van Weeren}, R.; {Hoeft}, M.
\newblock {XMM-Newton observations of the merging galaxy cluster CIZA
  J2242.8+5301}.
\newblock {\em Monthly Notices of the Royal Astronomical Society} {\bf 2013},
  {\em 429},~2617--2633,  \href{http://xxx.lanl.gov/abs/1201.1502}{{\normalfont
  [arXiv:astro-ph.CO/1201.1502]}}.
\newblock
  doi:{\changeurlcolor{black}\href{https://doi.org/10.1093/mnras/sts538}{\detokenize{10.1093/mnras/sts538}}}.

\bibitem[{Ogrean} \em{et~al.}(2014){Ogrean}, {Br{\"u}ggen}, {van Weeren},
  {R{\"o}ttgering}, {Simionescu}, {Hoeft}, and {Croston}]{Ogrean2014}
{Ogrean}, G.A.; {Br{\"u}ggen}, M.; {van Weeren}, R.; {R{\"o}ttgering}, H.;
  {Simionescu}, A.; {Hoeft}, M.; {Croston}, J.H.
\newblock {Multiple density discontinuities in the merging galaxy cluster CIZA
  J2242.8+5301}.
\newblock {\em Monthly Notices of the Royal Astronomical Society} {\bf 2014},
  {\em 440},~3416--3425,  \href{http://xxx.lanl.gov/abs/1403.5273}{{\normalfont
  [arXiv:astro-ph.GA/1403.5273]}}.
\newblock
  doi:{\changeurlcolor{black}\href{https://doi.org/10.1093/mnras/stu537}{\detokenize{10.1093/mnras/stu537}}}.

\bibitem[{van Weeren} \em{et~al.}(2010){van Weeren}, {R{\"o}ttgering},
  {Br{\"u}ggen}, and {Hoeft}]{vanWeeren2010}
{van Weeren}, R.J.; {R{\"o}ttgering}, H.J.A.; {Br{\"u}ggen}, M.; {Hoeft}, M.
\newblock {Particle Acceleration on Megaparsec Scales in a Merging Galaxy
  Cluster}.
\newblock {\em Science} {\bf 2010}, {\em 330},~347,
  \href{http://xxx.lanl.gov/abs/1010.4306}{{\normalfont
  [arXiv:astro-ph.CO/1010.4306]}}.
\newblock
  doi:{\changeurlcolor{black}\href{https://doi.org/10.1126/science.1194293}{\detokenize{10.1126/science.1194293}}}.

\bibitem[{ZuHone} \em{et~al.}(2021){ZuHone}, {Markevitch}, {Weinberger},
  {Nulsen}, and {Ehlert}]{ZuHone2021}
{ZuHone}, J.A.; {Markevitch}, M.; {Weinberger}, R.; {Nulsen}, P.; {Ehlert}, K.
\newblock {How Merger-driven Gas Motions in Galaxy Clusters Can Turn AGN
  Bubbles into Radio Relics}.
\newblock {\em Astrophysical Journal} {\bf 2021}, {\em 914},~73,
  \href{http://xxx.lanl.gov/abs/2012.02001}{{\normalfont
  [arXiv:astro-ph.HE/2012.02001]}}.
\newblock
  doi:{\changeurlcolor{black}\href{https://doi.org/10.3847/1538-4357/abf7bc}{\detokenize{10.3847/1538-4357/abf7bc}}}.

\bibitem[{Kulsrud} and {Pearce}(1969)]{Kulsrud1969}
{Kulsrud}, R.; {Pearce}, W.P.
\newblock {The Effect of Wave-Particle Interactions on the Propagation of
  Cosmic Rays}.
\newblock {\em Astrophysical Journal} {\bf 1969}, {\em 156},~445.
\newblock
  doi:{\changeurlcolor{black}\href{https://doi.org/10.1086/149981}{\detokenize{10.1086/149981}}}.

\bibitem[{Wiener} \em{et~al.}(2013){Wiener}, {Oh}, and {Guo}]{Wiener2013}
{Wiener}, J.; {Oh}, S.P.; {Guo}, F.
\newblock {Cosmic ray streaming in clusters of galaxies}.
\newblock {\em Monthly Notices of the Royal Astronomical Society} {\bf 2013},
  {\em 434},~2209--2228,  \href{http://xxx.lanl.gov/abs/1303.4746}{{\normalfont
  [arXiv:astro-ph.HE/1303.4746]}}.
\newblock
  doi:{\changeurlcolor{black}\href{https://doi.org/10.1093/mnras/stt1163}{\detokenize{10.1093/mnras/stt1163}}}.

\bibitem[{Springel}(2010)]{Springel2010}
{Springel}, V.
\newblock {E pur si muove: Galilean-invariant cosmological hydrodynamical
  simulations on a moving mesh}.
\newblock {\em Monthly Notices of the Royal Astronomical Society} {\bf 2010},
  {\em 401},~791--851,  \href{http://xxx.lanl.gov/abs/0901.4107}{{\normalfont
  [arXiv:astro-ph.CO/0901.4107]}}.
\newblock
  doi:{\changeurlcolor{black}\href{https://doi.org/10.1111/j.1365-2966.2009.15715.x}{\detokenize{10.1111/j.1365-2966.2009.15715.x}}}.

\bibitem[{Pakmor} and {Springel}(2013)]{Pakmor2013}
{Pakmor}, R.; {Springel}, V.
\newblock {Simulations of magnetic fields in isolated disc galaxies}.
\newblock {\em Monthly Notices of the Royal Astronomical Society} {\bf 2013},
  {\em 432},~176--193,  \href{http://xxx.lanl.gov/abs/1212.1452}{{\normalfont
  [arXiv:astro-ph.CO/1212.1452]}}.
\newblock
  doi:{\changeurlcolor{black}\href{https://doi.org/10.1093/mnras/stt428}{\detokenize{10.1093/mnras/stt428}}}.

\bibitem[{Marinacci} \em{et~al.}(2018){Marinacci}, {Vogelsberger}, {Pakmor},
  {Torrey}, {Springel}, {Hernquist}, {Nelson}, {Weinberger}, {Pillepich},
  {Naiman}, and {Genel}]{Marinacci2018}
{Marinacci}, F.; {Vogelsberger}, M.; {Pakmor}, R.; {Torrey}, P.; {Springel},
  V.; {Hernquist}, L.; {Nelson}, D.; {Weinberger}, R.; {Pillepich}, A.;
  {Naiman}, J.; {Genel}, S.
\newblock {First results from the IllustrisTNG simulations: radio haloes and
  magnetic fields}.
\newblock {\em Monthly Notices of the Royal Astronomical Society} {\bf 2018},
  {\em 480},~5113--5139,
  \href{http://xxx.lanl.gov/abs/1707.03396}{{\normalfont
  [arXiv:astro-ph.CO/1707.03396]}}.
\newblock
  doi:{\changeurlcolor{black}\href{https://doi.org/10.1093/mnras/sty2206}{\detokenize{10.1093/mnras/sty2206}}}.

\bibitem[{Ascasibar} and {Markevitch}(2006)]{Ascasibar2006}
{Ascasibar}, Y.; {Markevitch}, M.
\newblock {The Origin of Cold Fronts in the Cores of Relaxed Galaxy Clusters}.
\newblock {\em Astrophysical Journal} {\bf 2006}, {\em 650},~102--127,
  \href{http://xxx.lanl.gov/abs/astro-ph/0603246}{{\normalfont
  [arXiv:astro-ph/astro-ph/0603246]}}.
\newblock
  doi:{\changeurlcolor{black}\href{https://doi.org/10.1086/506508}{\detokenize{10.1086/506508}}}.

\bibitem[{ZuHone} \em{et~al.}(2010){ZuHone}, {Markevitch}, and
  {Johnson}]{ZuHone2010}
{ZuHone}, J.A.; {Markevitch}, M.; {Johnson}, R.E.
\newblock {Stirring Up the Pot: Can Cooling Flows in Galaxy Clusters be
  Quenched by Gas Sloshing?}
\newblock {\em Astrophysical Journal} {\bf 2010}, {\em 717},~908--928,
  \href{http://xxx.lanl.gov/abs/0912.0237}{{\normalfont
  [arXiv:astro-ph.CO/0912.0237]}}.
\newblock
  doi:{\changeurlcolor{black}\href{https://doi.org/10.1088/0004-637X/717/2/908}{\detokenize{10.1088/0004-637X/717/2/908}}}.

\bibitem[{ZuHone} \em{et~al.}(2016){ZuHone}, {Miller}, {Simionescu}, and
  {Bautz}]{ZuHone2016}
{ZuHone}, J.A.; {Miller}, E.D.; {Simionescu}, A.; {Bautz}, M.W.
\newblock {Simulating Astro-H Observations of Sloshing Gas Motions in the Cores
  of Galaxy Clusters}.
\newblock {\em Astrophysical Journal} {\bf 2016}, {\em 821},~6,
  \href{http://xxx.lanl.gov/abs/1508.04426}{{\normalfont
  [arXiv:astro-ph.HE/1508.04426]}}.
\newblock
  doi:{\changeurlcolor{black}\href{https://doi.org/10.3847/0004-637X/821/1/6}{\detokenize{10.3847/0004-637X/821/1/6}}}.

\bibitem[{ZuHone} \em{et~al.}(2018){ZuHone}, {Miller}, {Bulbul}, and
  {Zhuravleva}]{ZuHone2018}
{ZuHone}, J.A.; {Miller}, E.D.; {Bulbul}, E.; {Zhuravleva}, I.
\newblock {What Do the Hitomi Observations Tell Us About the Turbulent
  Velocities in the Perseus Cluster? Probing the Velocity Field with Mock
  Observations}.
\newblock {\em Astrophysical Journal} {\bf 2018}, {\em 853},~180,
  \href{http://xxx.lanl.gov/abs/1708.07206}{{\normalfont
  [arXiv:astro-ph.HE/1708.07206]}}.
\newblock
  doi:{\changeurlcolor{black}\href{https://doi.org/10.3847/1538-4357/aaa4b3}{\detokenize{10.3847/1538-4357/aaa4b3}}}.

\bibitem[{ZuHone} \em{et~al.}(2019){ZuHone}, {Zavala}, and
  {Vogelsberger}]{ZuHone2019}
{ZuHone}, J.A.; {Zavala}, J.; {Vogelsberger}, M.
\newblock {Sloshing of Galaxy Cluster Core Plasma in the Presence of
  Self-interacting Dark Matter}.
\newblock {\em Astrophysical Journal} {\bf 2019}, {\em 882},~119,
  \href{http://xxx.lanl.gov/abs/1901.11140}{{\normalfont
  [arXiv:astro-ph.CO/1901.11140]}}.
\newblock
  doi:{\changeurlcolor{black}\href{https://doi.org/10.3847/1538-4357/ab321d}{\detokenize{10.3847/1538-4357/ab321d}}}.

\bibitem[{Brzycki} and {ZuHone}(2019)]{Brzycki2019}
{Brzycki}, B.; {ZuHone}, J.
\newblock {A Parameter Space Exploration of Galaxy Cluster Mergers. II. Effects
  of Magnetic Fields}.
\newblock {\em Astrophysical Journal} {\bf 2019}, {\em 883},~118,
  \href{http://xxx.lanl.gov/abs/1904.10024}{{\normalfont
  [arXiv:astro-ph.CO/1904.10024]}}.
\newblock
  doi:{\changeurlcolor{black}\href{https://doi.org/10.3847/1538-4357/ab3983}{\detokenize{10.3847/1538-4357/ab3983}}}.

\bibitem[{Weinberger} \em{et~al.}(2017){Weinberger}, {Ehlert}, {Pfrommer},
  {Pakmor}, and {Springel}]{Weinberger2017}
{Weinberger}, R.; {Ehlert}, K.; {Pfrommer}, C.; {Pakmor}, R.; {Springel}, V.
\newblock {Simulating the interaction of jets with the intracluster medium}.
\newblock {\em Monthly Notices of the Royal Astronomical Society} {\bf 2017},
  {\em 470},~4530--4546,
  \href{http://xxx.lanl.gov/abs/1703.09223}{{\normalfont
  [arXiv:astro-ph.GA/1703.09223]}}.
\newblock
  doi:{\changeurlcolor{black}\href{https://doi.org/10.1093/mnras/stx1409}{\detokenize{10.1093/mnras/stx1409}}}.

\bibitem[{Pfrommer} \em{et~al.}(2017){Pfrommer}, {Pakmor}, {Schaal}, {Simpson},
  and {Springel}]{Pfrommer2017}
{Pfrommer}, C.; {Pakmor}, R.; {Schaal}, K.; {Simpson}, C.M.; {Springel}, V.
\newblock {Simulating cosmic ray physics on a moving mesh}.
\newblock {\em Monthly Notices of the Royal Astronomical Society} {\bf 2017},
  {\em 465},~4500--4529,
  \href{http://xxx.lanl.gov/abs/1604.07399}{{\normalfont
  [arXiv:astro-ph.GA/1604.07399]}}.
\newblock
  doi:{\changeurlcolor{black}\href{https://doi.org/10.1093/mnras/stw2941}{\detokenize{10.1093/mnras/stw2941}}}.

\bibitem[{Ehlert} \em{et~al.}(2018){Ehlert}, {Weinberger}, {Pfrommer},
  {Pakmor}, and {Springel}]{Ehlert2018}
{Ehlert}, K.; {Weinberger}, R.; {Pfrommer}, C.; {Pakmor}, R.; {Springel}, V.
\newblock {Simulations of the dynamics of magnetized jets and cosmic rays in
  galaxy clusters}.
\newblock {\em Monthly Notices of the Royal Astronomical Society} {\bf 2018},
  {\em 481},~2878--2900,
  \href{http://xxx.lanl.gov/abs/1806.05679}{{\normalfont
  [arXiv:astro-ph.CO/1806.05679]}}.
\newblock
  doi:{\changeurlcolor{black}\href{https://doi.org/10.1093/mnras/sty2397}{\detokenize{10.1093/mnras/sty2397}}}.

\bibitem[{Perucho} and {Mart{\'\i}}(2007)]{Perucho2007}
{Perucho}, M.; {Mart{\'\i}}, J.M.
\newblock {A numerical simulation of the evolution and fate of a Fanaroff-Riley
  type I jet. The case of 3C 31}.
\newblock {\em Monthly Notices of the Royal Astronomical Society} {\bf 2007},
  {\em 382},~526--542,  \href{http://xxx.lanl.gov/abs/0709.1784}{{\normalfont
  [arXiv:astro-ph/0709.1784]}}.
\newblock
  doi:{\changeurlcolor{black}\href{https://doi.org/10.1111/j.1365-2966.2007.12454.x}{\detokenize{10.1111/j.1365-2966.2007.12454.x}}}.

\bibitem[{Wiener} \em{et~al.}(2017){Wiener}, {Pfrommer}, and {Oh}]{Wiener2017}
{Wiener}, J.; {Pfrommer}, C.; {Oh}, S.P.
\newblock {Cosmic ray-driven galactic winds: streaming or diffusion?}
\newblock {\em Monthly Notices of the Royal Astronomical Society} {\bf 2017},
  {\em 467},~906--921,  \href{http://xxx.lanl.gov/abs/1608.02585}{{\normalfont
  [arXiv:astro-ph.GA/1608.02585]}}.
\newblock
  doi:{\changeurlcolor{black}\href{https://doi.org/10.1093/mnras/stx127}{\detokenize{10.1093/mnras/stx127}}}.

\bibitem[{Pakmor} \em{et~al.}(2016){Pakmor}, {Pfrommer}, {Simpson}, {Kannan},
  and {Springel}]{Pakmor2016}
{Pakmor}, R.; {Pfrommer}, C.; {Simpson}, C.M.; {Kannan}, R.; {Springel}, V.
\newblock {Semi-implicit anisotropic cosmic ray transport on an unstructured
  moving mesh}.
\newblock {\em Monthly Notices of the Royal Astronomical Society} {\bf 2016},
  {\em 462},~2603--2616,
  \href{http://xxx.lanl.gov/abs/1604.08587}{{\normalfont
  [arXiv:astro-ph.GA/1604.08587]}}.
\newblock
  doi:{\changeurlcolor{black}\href{https://doi.org/10.1093/mnras/stw1761}{\detokenize{10.1093/mnras/stw1761}}}.

\bibitem[{Desiati} and {Zweibel}(2014)]{Desiati2014}
{Desiati}, P.; {Zweibel}, E.G.
\newblock {The Transport of Cosmic Rays Across Magnetic Fieldlines}.
\newblock {\em Astrophysical Journal} {\bf 2014}, {\em 791},~51,
  \href{http://xxx.lanl.gov/abs/1402.1475}{{\normalfont
  [arXiv:astro-ph.HE/1402.1475]}}.
\newblock
  doi:{\changeurlcolor{black}\href{https://doi.org/10.1088/0004-637X/791/1/51}{\detokenize{10.1088/0004-637X/791/1/51}}}.

\bibitem[{Pfrommer}(2013)]{Pfrommer2013}
{Pfrommer}, C.
\newblock {Toward a Comprehensive Model for Feedback by Active Galactic Nuclei:
  New Insights from M87 Observations by LOFAR, Fermi, and H.E.S.S.}
\newblock {\em Astrophysical Journal} {\bf 2013}, {\em 779},~10,
  \href{http://xxx.lanl.gov/abs/1303.5443}{{\normalfont
  [arXiv:astro-ph.CO/1303.5443]}}.
\newblock
  doi:{\changeurlcolor{black}\href{https://doi.org/10.1088/0004-637X/779/1/10}{\detokenize{10.1088/0004-637X/779/1/10}}}.

\bibitem[{Kempski} and {Quataert}(2020)]{Kempski2020}
{Kempski}, P.; {Quataert}, E.
\newblock {Thermal instability of halo gas heated by streaming cosmic rays}.
\newblock {\em Monthly Notices of the Royal Astronomical Society} {\bf 2020},
  {\em 493},~1801--1817,
  \href{http://xxx.lanl.gov/abs/1908.10367}{{\normalfont
  [arXiv:astro-ph.GA/1908.10367]}}.
\newblock
  doi:{\changeurlcolor{black}\href{https://doi.org/10.1093/mnras/staa385}{\detokenize{10.1093/mnras/staa385}}}.

\bibitem[{ZuHone} \em{et~al.}(2018){ZuHone}, {Kowalik}, {{\"O}hman}, {Lau}, and
  {Nagai}]{ZuHoneGCMC}
{ZuHone}, J.A.; {Kowalik}, K.; {{\"O}hman}, E.; {Lau}, E.; {Nagai}, D.
\newblock {The Galaxy Cluster Merger Catalog: An Online Repository of Mock
  Observations from Simulated Galaxy Cluster Mergers}.
\newblock {\em Astrophysical Journal Supplement Series} {\bf 2018}, {\em
  234},~4.
\newblock
  doi:{\changeurlcolor{black}\href{https://doi.org/10.3847/1538-4365/aa99db}{\detokenize{10.3847/1538-4365/aa99db}}}.

\bibitem[{Astropy Collaboration} \em{et~al.}(2013){Astropy Collaboration},
  {Robitaille}, {Tollerud}, {Greenfield}, {Droettboom}, {Bray}, {Aldcroft},
  {Davis}, {Ginsburg}, {Price-Whelan}, {Kerzendorf}, {Conley}, {Crighton},
  {Barbary}, {Muna}, {Ferguson}, {Grollier}, {Parikh}, {Nair}, {Unther},
  {Deil}, {Woillez}, {Conseil}, {Kramer}, {Turner}, {Singer}, {Fox}, {Weaver},
  {Zabalza}, {Edwards}, {Azalee Bostroem}, {Burke}, {Casey}, {Crawford},
  {Dencheva}, {Ely}, {Jenness}, {Labrie}, {Lim}, {Pierfederici}, {Pontzen},
  {Ptak}, {Refsdal}, {Servillat}, and {Streicher}]{AstroPy2013}
{Astropy Collaboration}.; {Robitaille}, T.P.; {Tollerud}, E.J.; {Greenfield},
  P.; {Droettboom}, M.; {Bray}, E.; {Aldcroft}, T.; {Davis}, M.; {Ginsburg},
  A.; {Price-Whelan}, A.M.; {Kerzendorf}, W.E.; {Conley}, A.; {Crighton}, N.;
  {Barbary}, K.; {Muna}, D.; {Ferguson}, H.; {Grollier}, F.; {Parikh}, M.M.;
  {Nair}, P.H.; {Unther}, H.M.; {Deil}, C.; {Woillez}, J.; {Conseil}, S.;
  {Kramer}, R.; {Turner}, J.E.H.; {Singer}, L.; {Fox}, R.; {Weaver}, B.A.;
  {Zabalza}, V.; {Edwards}, Z.I.; {Azalee Bostroem}, K.; {Burke}, D.J.;
  {Casey}, A.R.; {Crawford}, S.M.; {Dencheva}, N.; {Ely}, J.; {Jenness}, T.;
  {Labrie}, K.; {Lim}, P.L.; {Pierfederici}, F.; {Pontzen}, A.; {Ptak}, A.;
  {Refsdal}, B.; {Servillat}, M.; {Streicher}, O.
\newblock {Astropy: A community Python package for astronomy}.
\newblock {\em Astronomy \& Astrophysics} {\bf 2013}, {\em 558},~A33,
  \href{http://xxx.lanl.gov/abs/1307.6212}{{\normalfont
  [arXiv:astro-ph.IM/1307.6212]}}.
\newblock
  doi:{\changeurlcolor{black}\href{https://doi.org/10.1051/0004-6361/201322068}{\detokenize{10.1051/0004-6361/201322068}}}.

\bibitem[Hunter(2007)]{Hunter2007}
Hunter, J.D.
\newblock Matplotlib: A 2D graphics environment.
\newblock {\em Computing in Science \& Engineering} {\bf 2007}, {\em
  9},~90--95.
\newblock
  doi:{\changeurlcolor{black}\href{https://doi.org/10.1109/MCSE.2007.55}{\detokenize{10.1109/MCSE.2007.55}}}.

\bibitem[Harris \em{et~al.}(2020)Harris, Millman, van~der Walt, Gommers,
  Virtanen, Cournapeau, Wieser, Taylor, Berg, Smith, Kern, Picus, Hoyer, van
  Kerkwijk, Brett, Haldane, del R{\'{i}}o, Wiebe, Peterson,
  G{\'{e}}rard-Marchant, Sheppard, Reddy, Weckesser, Abbasi, Gohlke, and
  Oliphant]{Harris2020}
Harris, C.R.; Millman, K.J.; van~der Walt, S.J.; Gommers, R.; Virtanen, P.;
  Cournapeau, D.; Wieser, E.; Taylor, J.; Berg, S.; Smith, N.J.; Kern, R.;
  Picus, M.; Hoyer, S.; van Kerkwijk, M.H.; Brett, M.; Haldane, A.; del
  R{\'{i}}o, J.F.; Wiebe, M.; Peterson, P.; G{\'{e}}rard-Marchant, P.;
  Sheppard, K.; Reddy, T.; Weckesser, W.; Abbasi, H.; Gohlke, C.; Oliphant,
  T.E.
\newblock Array programming with {NumPy}.
\newblock {\em Nature} {\bf 2020}, {\em 585},~357--362.
\newblock
  doi:{\changeurlcolor{black}\href{https://doi.org/10.1038/s41586-020-2649-2}{\detokenize{10.1038/s41586-020-2649-2}}}.

\bibitem[{Turk} \em{et~al.}(2011){Turk}, {Smith}, {Oishi}, {Skory}, {Skillman},
  {Abel}, and {Norman}]{Turk2011}
{Turk}, M.J.; {Smith}, B.D.; {Oishi}, J.S.; {Skory}, S.; {Skillman}, S.W.;
  {Abel}, T.; {Norman}, M.L.
\newblock {yt: A Multi-code Analysis Toolkit for Astrophysical Simulation
  Data}.
\newblock {\em Astrophysical Journals} {\bf 2011}, {\em 192},~9,
  \href{http://xxx.lanl.gov/abs/1011.3514}{{\normalfont
  [arXiv:astro-ph.IM/1011.3514]}}.
\newblock
  doi:{\changeurlcolor{black}\href{https://doi.org/10.1088/0067-0049/192/1/9}{\detokenize{10.1088/0067-0049/192/1/9}}}.

\end{thebibliography}

%

%


\end{document}